\documentclass[letterpaper, 10 pt, conference]{ieeeconf}  % Comment this line out if you need a4paper

\IEEEoverridecommandlockouts                              % This command is only needed if 
\usepackage{color}
\usepackage{amsmath}
\usepackage{amssymb}
\usepackage{algorithmic,algorithm}
\usepackage{tikz} \usetikzlibrary{arrows}
\newcommand{\ignore}[1]{}

\newtheorem{thm}{Theorem}
\newtheorem{lem}[thm]{Lemma}
\def\Es{\mathcal{E}} % Edge set
\def\Ns{\mathcal{N}} % Node set
\def\Ps{\mathcal{P}} % Participating node set
\def\NPs{\bar{\mathcal{P}}} % Non-participating node set
\def\ie{\emph{i.e.,~}}
\def\xibf{\boldsymbol{\xi}}

\def\xbf{\mathbf{x}}
\def\Fbf{\mathbf{F}}
\def\Abf{\mathbf{A}}
\def\Lbf{\mathbf{L}}
\def\Ibf{\mathbf{I}}

\def\Qbf{\mathbf{Q}}

\def\ybf{\mathbf{y}}
\def\Pbf{\mathbf{P}}
\def\Ybf{\mathbf{Y}}
\def\Mbf{\mathbf{M}}

\def\ibf{\mathbf{i}}

\title{\LARGE \bf
Bridge Consensus: Ignoring Initial Inessentials
}
\author{David W. Casbeer$^{1}$, Yongcan Cao$^{1}$, Eloy Garcia$^{3}$ and  Dejan Milutinovic$^{4}$% <-this % stops a space
\thanks{$^{1}$D. Casbeer \& Y. Cao are with AFRL's Control Science Center of Excellence, Wright-Patterson AFB, OH,  {\tt\small david.casbeer@us.af.mil} \& {\tt\small yongcan.cao@gmail.com}}%
\thanks{$^{2}$E. Garcia is a contractor (Infoscitex Corp.) with AFRL's Control Science Center of Excellence, Wright-Patterson AFB, OH, {\tt\small elgarcia@infoscitex.com}}%
\thanks{$^{3}$D. Milutinovic is with the Department of Electrical Engineering, UC Santa Cruz, {\tt\small dejan@soe.ucsc.edu}}%
}

\begin{document}

\maketitle

\begin{abstract}
In this paper, the problem of bridge consensus is presented and solved.  Bridge consensus consists of a network of nodes, some of whom are participating and others are non-participating.  The objective is for all the agents to reach average consensus of the participating nodes initial values in a distributed and scalable manner.  To do this, the nodes must use the network connections of the non-participating nodes, which act as bridges for information and ignore the initial values of the non-participating nodes.  The solution to this problem is made by merging the ideas from estimation theory and consensus theory.  By considering the participating nodes has having equal information and the non-participating nodes as having no information, the nodes initial values are transformed into information space.  Two consensus filters are run in parallel on the information state and information matrix.  Conditions ensuring that the product of the inverse information matrix and the information state of each agent reaches average consensus of the participating agents' initial values is given.
\end{abstract}

%%%%%%%%%%%%%%%%%%%%%%%%%%%%%%%%%%%%%%%%%%%%%%%%%%%%%%%%%%%%%%%%%%%%%%%%%%%%%%%%%%%%%%%%%%%%%%%%%%%%%%%%%
\section{Introduction}
%%%%%%%%%%%%%%%%%%%%%%%%%%%%%%%%%%%%%%%%%%%%%%%%%%%%%%%%%%%%%%%%%%%%%%%%%%%%%%%%%%%%%%%%%%%%%%%%%%%%%%%%%
Suppose a scenario, where each node in a wireless sensor network needs to estimate the state of the object/process being observed.  Furthermore, suppose that a number of sensors lose their sensor capability or due to limited sensor capabilities they are unable to observe the object/process.  In order for each sensor to maintain a local estimate, a distributed and scalable data fusion mechanism becomes necessary.  The sensors with no observation are still able to communicate with their neighbors and are most likely necessary to keep the network connected.  

This paper presents a distributed and scalable solution to this problem, which is aptly called \emph{bridge consensus}, since the sensors without observations must act as ``bridges'' to relay information, without themselves having an opinion/information to contribute to the data fusion process.  First a simple example is shown to illustrate the difficulty of this problem.  Then in Sections \ref{sec:max-likelihood} and \ref{sec:consFilt} we will briefly present the relevant ideas from estimation and consensus theory that will be combined together in Section  \ref{sec:ignore} to solve the bridge consensus problem.  The paper is concluded with a simulation to verify the results in this paper and a few closing statements.

%%%%
\subsection{Example}
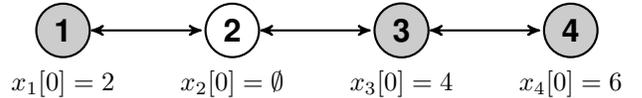
\begin{figure}[htb]\centering
\begin{tikzpicture}[<->,>=stealth',shorten >=1pt,auto,node distance=2.25cm,
  	thick,main node/.style={circle,fill=black!20,draw,font=\sffamily\large\bfseries}]
  	\node[main node] (1) {1};
  	\node[circle, draw, font=\sffamily\large\bfseries] (2) [right of=1] {2};
  	\node[main node] (3) [right of=2] {3};
  	\node[main node] (4) [right of=3] {4};
	\path[every node/.style={font=\sffamily\small}]
	   (1) edge node [left] {} (2)
	   (2) edge node [left] {} (3)
	   (3) edge node [left] {} (4);
	\node (t1) at (1) [below = .4cm] {$x_{1}[0] = 2$};
	\node (t2) at (2) [below = .4cm] {$x_{2}[0] = \emptyset$};
	\node (t3) at (3) [below = .4cm] {$x_{3}[0] = 4$};
	\node (t4) at (4) [below = .4cm] {$x_{4}[0] = 6$};
\end{tikzpicture}
\caption{An example graph where nodes 1, 3, and 4 are participating and node 2 is not participating}
\label{fig:example}
\end{figure}

Suppose four agents in a undirected graph, as pictured in Fig. \ref{fig:example}.  Agents 1, 3, and 4 have initial values as depicted in the figure. However, agent 2 does not have an initial opinion and, therefore, does not have an initial value to contribute.  The objective of the four agents is to come to \emph{average} consensus on the initial values of the participating agents (\ie 1, 3, and 4), while ignoring the non-value held initially by agent 2.  Furthermore, we want the method to be scalable with the number of agents and degree of the graph.  It is clear, that without the network connection that agent 2 provides, it would be impossible to reach a consensus.  The next two paragraphs discuss two options that seem to be reasonable approaches, but will not work.  They are included to help better understand the difficulty of this problem.

For discussion's sake, let us consider a discrete-time implementation where each agent takes the average of its neighbors.  Suppose that agent 2 seeks to ``initialize'' its value with some combination of its neighbors values.  After the first time step, the states of each agent would become: $x_{1} = 2$, $x_{2} = 3$, $x_{3} = 5$, $x_{4} = 5$; the average of which is 3.75.  From this example, we see that a simple initialization of the non-participating nodes using their neighbors values will not work, because it biases the average towards the values held by the non-participating nodes' neighbors.  Perhaps there is a way to perform a sophisticated initialization step where the neighbors and the non-participating nodes adjust their values, but at first glance a solution is not apparent.  

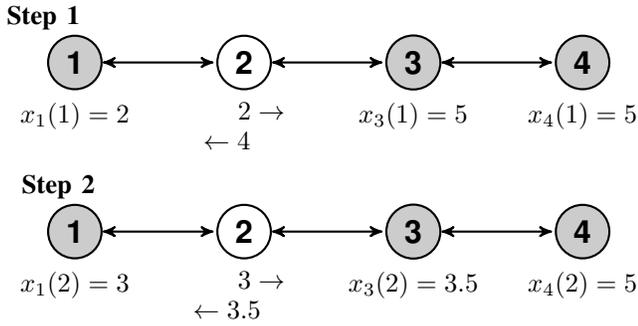
\begin{figure}[h]\centering
\begin{tikzpicture}[<->,>=stealth',shorten >=1pt,auto,node distance=2.25cm,
  	thick,main node/.style={circle,fill=black!20,draw,font=\sffamily\large\bfseries}]
  	\node[main node] (1) {1};
  	\node[circle, draw, font=\sffamily\large\bfseries] (2) [right of=1] {2};
  	\node[main node] (3) [right of=2] {3};
  	\node[main node] (4) [right of=3] {4};
	\path[every node/.style={font=\sffamily\small}]
	   (1) edge node [left] {} (2)
	   (2) edge node [left] {} (3)
	   (3) edge node [left] {} (4);
	\node (time1) at (1) [above = .6cm, left = -.2cm] {\textbf{Step 1}};
	\node (t1) at (1) [below = .4cm] {$x_{1}(1) = 2$};
	\node (t2) at (2) [below = .4cm] {$ \hspace{13pt} 2 \rightarrow$};
	\node (t2b) at (2) [below = .8cm] {$\leftarrow 4  \hspace{13pt} $};	
	\node (t3) at (3) [below = .4cm] {$x_{3}(1) = 5$};
	\node (t4) at (4) [below = .4cm] {$x_{4}(1) = 5$};
	%%%%
  	\node[main node] (b1) at (1) [below of = 1] {1};
  	\node[circle, draw, font=\sffamily\large\bfseries] (b2) [right of=b1] {2};
  	\node[main node] (b3) [right of=b2] {3};
  	\node[main node] (b4) [right of=b3] {4};
	\path[every node/.style={font=\sffamily\small}]
	   (b1) edge node [left] {} (b2)
	   (b2) edge node [left] {} (b3)
	   (b3) edge node [left] {} (b4);
	\node (time2) at (b1) [above = .6cm, left = -.4cm] {\textbf{Step 2}};
	\node (bt1) at (b1) [below = .4cm] {$x_{1}(2) = 3$};
	\node (bt2) at (b2) [below = .4cm] {$ \hspace{13pt} 3 \rightarrow$};
	\node (bt2b) at (b2) [below = .8cm] {$\leftarrow 3.5  \hspace{13pt} $};	
	\node (bt3) at (b3) [below = .4cm] {$x_{3}(2) = 3.5$};
	\node (bt4) at (b4) [below = .4cm] {$x_{4}(2) = 5$};	
\end{tikzpicture}
\caption{Example where the non-participating node, agent 2, acts as a relay.}
\label{fig:example-relay}
\end{figure}

Another possible avenue to attack this problem is to think of the non-participating node as a relay of information.  Here agent 2 would store it's neighbors values, and pass those along to its neighbors.  Figure \ref{fig:example-relay} shows one possibile scenario of how this could happen.  At time 1, agent 2 stores the information it receives from its neighbors.  At time 2, agent 2 has passed this information to its neighbors, while simultaneously storing the information it receives from it's neighbors.  With such a scheme, the participating agents would converge to some value that is a combination of their initial values, and agent 2 would also know this value.  At time 1, the average of the participating agents is 4, which is the average of the initial conditions.  However, at time 2, the average of the participating agents is 3.8.  The delay of information caused the system to lose the average.  Of course, for this simple example, if agent 2 were able to instantaneously relay its neighbors' values then there would be no problem.  However, as the degree of the non-participating nodes increases or as the number of connected non-participating nodes increases.  The problem of exactly how to relay or route the information in a decentralized manner comes into question.  

In summary, it is not clear how distributed averaging algorithms \cite{Jadbabaie2003} or gossip algorithms \cite{Boyd2006aa,Aysal2009aa} can be adapted to solve the average consensus problem with the specific constraint that a subset of the nodes do not have an initial value to contribute, yet, they are necessary to maintain connectivity of the communication graph.  Obviously, the application of gossip algorithms needs to respect the physical limitations of the graph.  For instance, in the example shown above, node 1 is never able to establish direct communication with either node 3 or node 4 and the link that node 2 offers is essential to converge to a global common value.

%%%%%%%%%%%%%%%%%%%%%%%%%%%%%%%%%%%%%%%%%%%%%%%%%%%%%%%%%%%%%%%%%%%%%%%%%%%%%%%%%%%%%%%%%%%%%%%%%%%%%%%%%
\section{Bridge Consensus}
\label{sec:bridge-consensus}

%\subsection{Problem Formulation}
%\label{sec:problem}
%%%%%%%%%%%%%%%%%%%%%%%%%%%%%%%%%%%%%%%%%%%%%%%%%%%%%%%%%%%%%%%%%%%%%%%%%%%%%%%%%%%%%%%%%%%%%%%%%%%%%%%%%

%In the bridge consensus problem, there exists of set of nodes, denoted by $\Ns$.  The nodes are connected by edges, and the edge $(i,j)$ is in the set, $\Es$, of edges if agent $i$ can send information to agent $j$.  Node $i \in \Ns$ is an incoming neighbor of agent $j \in \Ns$ if the edge $(i,j) \in \Es$.  The set of all incoming neighbors for agent $j$ is denote by $\Ns_{j}$.  Let $|\Ns|$ denote the number of elements in the set $\Ns$.  

Given and set of nodes, denoted by $\Ns$, the \emph{bridge consensus} problem is for all of the nodes to reach average consensus on the initial values of a subset of these nodes, in a distributed and scalable fashion.  The subset of nodes whose initial values are to contribute to the final average are denoted by the \emph{participating} node set $\Ps \subset \Ns$.  The rest of the nodes are called \emph{non-participating}, and this subset is denoted by $\NPs = \Ns \setminus \Ps$.  It is important that the non-participating nodes share their communication links so that the entire set of nodes can reach a consensus on the average value of the participating nodes.

To solve the bridge consensus problem, we combine the ideas from estimation theory and consensus literature.  First, Section \ref{sec:max-likelihood} briefly presents the maximum-likelihood mean estimate of the of independent normally distributed random variables.  Section \ref{sec:consFilt} follows with the necessary formalisms in consensus theory.  Then, in Section \ref{sec:ignore}, these two ideas are combined to solve the bridge consensus problem.

\subsection{Maximum-likelihood Mean Estimate}
\label{sec:max-likelihood}
Suppose that we have $n$ independent normally distributed random samples whose distributions are given by 
\begin{equation}
x_{i} \sim \mathcal{N}(\mu, R_{i}), \; \text{for} \; i=1,\dots,n,
\end{equation}
that is, each random sample has the same mean, but the variances differ.  It is well known that the maximum likelihood estimate of the mean, given these samples is given by 
\begin{equation}
\label{eq:ml}
\hat{\mu} = \left(\sum_{i=1}^{n}R_{i}^{-1}\right)^{-1} \sum_{i=1}^{n} R_{i}^{-1}x_{i}.
\end{equation}
Rewriting Equation \eqref{eq:ml} in the information form yields
\begin{equation}
\label{eq:ml-info}
\hat{\mu} = \left(\sum_{i=1}^{n}Y_{i}\right)^{-1} \sum_{i=1}^{n} y_{i},
\end{equation}
where the \emph{information matrix} and \emph{information state} are given, respectively, by $Y_{i} \triangleq R_{i}^{-1}$ and $y_{i} \triangleq Y_{i}x_{i}$.  Equation \eqref{eq:ml} can be equivalently written as
\begin{equation}
\label{eq:ml-average}
\hat{\mu} = \left(\frac{1}{n}\sum_{i=1}^{n}Y_{i}\right)^{-1} \frac{1}{n}\sum_{i=1}^{n} y_{i},
\end{equation}
which shows that if one were able to compute the average of the information state and information matrix, then the maximum likelihood estimate of the mean would be easily computed.

\subsection{Consensus Filter}
\label{sec:consFilt}

Consensus algorithms are decentralized methods for a team of agents to
agree on specific \emph{consensus states}.  In a
consensus filter, each agent exchanges information with neighboring
agents and not the entire team.  Over time the agents reach an
agreement (or consensus) concerning the consensus state
\cite{RenBeardAtkins2007, Olfati-SaberFaxMurray2007}.  Furthermore, \emph{average}
consensus occurs when the final consensus state is the average of
the initial values \cite{Olfati-Saber2004,Kingston2006}.

Before presenting the consensus algorithm used in this paper, some
graph theory terminology is needed. At any discrete-time instant $\tau$,
the communication topology between $n$ agents can be described by the
graph $G[\tau] = (\Ns,\Es[\tau])$ where $\Ns = \{1,\dots,n\}$ is the vertex set
and $\Es[\tau]\subseteq \Ns\times \Ns$ is the edge set.  The $ij^{th}$ element
of the \emph{adjacency matrix} $\Abf[\tau]$ of graph $G[\tau]$ is $A_{ij}[\tau] = 1$
if $i \neq j$ and the edge $(j,i) \in \Es[\tau]$, otherwise $A_{ij}[\tau] =
0$.  From the adjacency matrix one can construct the graph's Laplacian
matrix $\Lbf[\tau]$ as
\begin{align}
  \label{eq:Laplacian}
  L_{ij}[\tau] =
  \begin{cases}
    -A_{ij}[\tau] & \text{if $i \neq j$}, \\
    \sum_{j=1,j\neq i}^n A_{ij}[\tau] & \text{if $i = j$} \;.
  \end{cases}
\end{align}

In the consensus algorithm employed in this paper, each agent in the
network maintains a local copy of the consensus state $\xi_i \in
\mathbb{R}^m$. Each agent $i$ updates $\xi_i$ using its neighbors'
consensus states according to the rule
\begin{align}
  \label{eq:cLUpdate}
  %\xi_i[\tau+1] = \xi_i[\tau] - \frac{1}{d_\tau}\sum_{i=1}^n A_{ij}[\tau] (\xi_i[\tau] - \xi_j[\tau])
\xi_i[\tau+1] = \xi_i[\tau] - \frac{1}{d_\tau}\sum_{j=1}^n A_{ij}[\tau] (\xi_i[\tau] - \xi_j[\tau])
\end{align}
where $d_\tau \in [d_{max,\tau},\infty)$ and $d_{max,\tau}$ denotes the maximal
degree of $G[\tau]$.  The consensus protocol given in Eq.~\eqref{eq:cLUpdate} was chosen by the need for discrete-time average
consensus~\cite{Kingston2006}.  A nice distributed scheme to choose the
weights $d_\tau$ is to use \emph{Metropolis weights}~\cite{Boyd2004}.

After arranging the local information states into the vector
$\xibf[\tau] = [\xi_1^T[\tau], \dots, \xi_n^T[\tau]]^T$, the update can be
written as
\begin{align}
  \label{eq:cGUpdate}
  \xibf[\tau+1] &= (\Psi[\tau]\otimes\Ibf) \xibf[\tau]% = (\Ibf_{n} - \frac{1}{d_k}
  %\Lbf[k])\otimes \Ibf_{m} \xibf[k]
\end{align}
where,\vspace{-.25cm}
\begin{align}
  \label{eq:PsiDef}
  \Psi[\tau] &= \Ibf - \frac{1}{d_\tau} \Lbf[\tau] \;,
\end{align}
$\Ibf$ is the appropriate size identity matrix, and $\otimes$ denotes
the matrix Kronecker product.  Note that $\Psi[\tau]$ defined in this
manner is a stochastic matrix.
% \ignore{Journal show $\Psi$ is doubly stochastic.}

\begin{lem}~\cite{Kingston2006}\label{lem:av_consensus}
For a team of $n$ agents whose states satisfy~\eqref{eq:cLUpdate}, $\xi_i[\tau]\to\frac{1}{n} \sum_{i=1}^n \xi_i[0]$ as $\tau\to\infty$ if
\begin{itemize}
\item[1.] $G[\tau]$ is balanced for every $\tau$;
\item[2.] For any $\tau_0$, there exists $T$ such that the union of the graphs over the time interval $[\tau_0,\tau_0+T]$ is strongly connected. 
\end{itemize}
\end{lem}

%%%%%%%%%%%%%%%%%%%%%%%%%%%%%%%%%%%%%%%%%%%%%%%%%%%%%%%%%%%%%%%%%%%%%%%%%%%%%%%%%%%%%%%%%%%%%%%%%%%%%%%%%
\section{Bridge Consensus Solution}
\label{sec:ignore}

Consider the participating nodes in the bridge consensus problem.  These nodes can be thought of as being equally important or containing the same information content, while the non-participating nodes carry zero importance or information.  Using the insights from Sections \ref{sec:max-likelihood} and \ref{sec:consFilt}, suppose that we initialize an information state and information matrix for each agent as follows,
\begin{equation}
\label{eq:info-mat}
Y_{i}[0] = \begin{cases}
0& \text{if $i \in \NPs$}, \\
C& \text{if $i \in \Ps$, and}
\end{cases}
\end{equation}
\begin{equation}
\label{eq:info-state}
y_{i}[0] = \begin{cases}
0& \text{if $i \in \NPs$}, \\
Y_{i}[0]x_{i}[0]& \text{if $i \in \Ps$},
\end{cases}
\end{equation}
where $C$ is any positive-definite matrix.

%%% TODO: Write this in a theorem format
Let the network of agents implement two consensus filters according to Eq. \eqref{eq:cLUpdate}, one for the information matrix \eqref{eq:info-mat} and one for the information state \eqref{eq:info-state}.  Assuming that the conditions for average consensus are met (\ie Lemma \ref{lem:av_consensus}), then the agent's local values will converge asymptotically towards
\begin{eqnarray}
Y_{i}[\infty]  &=& \frac{1}{N} \sum_{i=1}^{N}Y_{i}[0], \; \forall i \in \Ns,\\
y_{i}[\infty] &=& \frac{1}{N} \sum_{i=1}^{N} y_{i}[0], \; \forall i \in \Ns,
\end{eqnarray}
Dropping the node subscript, and using this converged value to compute the maximum-likelihood mean estimate (Equation \eqref{eq:ml-average}) yields 
\begin{eqnarray}
\mu[\infty] &\triangleq& \left(Y_{i}[\infty]\right)^{-1} y_{i}[\infty]\\
&=& \left(\frac{1}{N}\sum_{i=1}^{N}Y_{i}[0]\right)^{-1} \frac{1}{N}\sum_{i=1}^{N} y_{i}[0] \\
&=& \left(\sum_{i\in\Ps}C\right)^{-1} \sum_{i\in\Ps} C x_{i}[0] \\
&=& \frac{1}{|\Ps|} C^{-1} C \sum_{i\in\Ps} x_{i}[0] \\
&=& \frac{1}{|\Ps|} \sum_{i\in\Ps} x_{i}[0]
\end{eqnarray}
which is the average of the participating nodes' initial values.

\section{Simulation}
\label{sec:results}

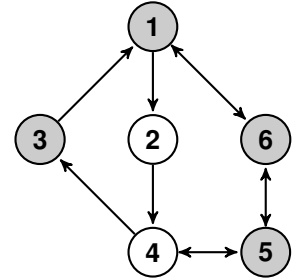
\begin{figure}[htbp]\centering
\begin{tikzpicture}[>=stealth',shorten >=1pt,auto,node distance=1.5cm,
  	thick,main node/.style={circle,fill=black!20,draw,font=\sffamily\bfseries}]
  	\node[main node] (1) {1};
  	\node[circle, draw, font=\sffamily\bfseries] (2) [below of=1] {2};
  	\node[main node] (3) [left of=2] {3};
  	\node[circle, draw, font=\sffamily\bfseries] (4) [below of=2] {4};
	\node[main node] (5) [right of=4] {5};
	\node[main node] (6) [right of =2] {6};
	\path[->] (1) edge node {} (2);
	\path[->] (2) edge node {} (4);
	\path[->] (3) edge node {} (1);
	\path[->] (4) edge node {} (3);
	\path[<->] (4) edge node {} (5);
	\path[<->] (5) edge node {} (6);	
	\path[<->] (1) edge node {} (6);	
%	\path[every node/.style={font=\sffamily\small}]
%	   (1) edge node [left] {} (2)
%	   (2) edge node [left] {} (3)
%	   (3) edge node [left] {} (4);
\end{tikzpicture}
\caption{In this example, the nodes desire to reach average consensus from the participating (gray) nodes (1, 3, 6,and 5) and where the white nodes 2 and 4 are not-participating.  Without node 2 and 4's communication channels, average consensus is impossible, since node 3 would have no incoming edges.}
\label{fig:simulation-graph}
\end{figure}

\begin{figure}[htbp]
\begin{center}
\includegraphics[trim={1.7cm 8.5cm 1.7cm 8.8cm},clip,width=.5\textwidth]{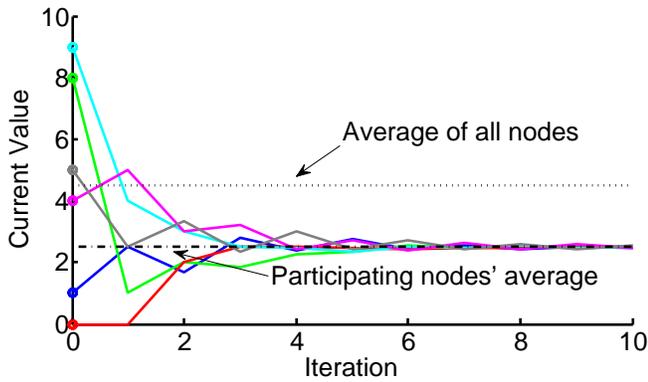}
\caption{Six agents running bridge consensus.  Participating agents' initial values are 0, 1, 4, and 5.  Non-participating agents' initial values are 9 and 10.  The agent converge to a value of 2.5, which is the average of the participating agents' initial values.}
\label{fig:simulation}
\end{center}
\end{figure}

To verify the results of the paper, a simple example is presented.  In this example, 6 nodes are connected in a graph as depicted in Figure \ref{fig:simulation-graph}.  The white nodes, 2 and 4, are not-participating, while the gray nodes are participating.  The initial values at each node are given by:
\begin{align*}
x_{1}[0] &= 1 
&x_{4}[0] &= 9 \\
x_{2}[0] &= 10
&x_{5}[0] &= 4 \\
x_{3}[0] &= 0 
&x_{6}[0] &= 5 
\end{align*}
It can be seen in Figure \ref{fig:simulation} that the nodes converge to the average of the participating nodes, namely $\frac{1}{4}\sum_{i\in \Ps} x_{i}[0] = \frac{1}{4}(1 + 0 + 9 + 10) = 2.5$.  The nodes ignored the initial values of the participating nodes 2 and 4; the average if all the nodes were included is 4.5.

%%%%%%%%%%%%%%%%%%%%%%%%%%%%%%%%%%%%%%%%%%%%%%%%%%%%%%%%%%%%%%%%%%%%%%%%%%%%%%%%%%%%%%%%%%%%%%%%%%%%%%%%%
\section{Conclusion}
\label{sec:conclusion}
In this paper, the problem of bridge consensus was presented and solved.  Bridge consensus consists of a network of nodes, some of whom are participating and others are not-participating.  The objective is for all the agents to reach average consensus of the participating nodes initial values in a distributed and scalable manner.  To solve this problem, ideas from estimation theory and consensus theory were merged.  By considering the participating nodes has having equal information and the non-participating nodes as having no information, the nodes initial values are transformed into information space.  Two consensus filters are run in parallel on the information state and information matrix.  Conditions ensuring that the product of the inverse information matrix and the information state of each agent reaches average consensus of the participating agents' initial values was given.

%%%%%%%%%%%%%%%%%%%%%%%%%%%%%%%%%%%%%%%%%%%%%%%%%%%%%%%%%%%%%%%%%%%%%%%%%%%%%%%%%%%%%%%%%%%%%%%%%%%%%%%%%
\bibliographystyle{plain}
\bibliography{KalmanConsensus}

\end{document}